# Identifying the Bottleneck for Heat Transport in Metal-Organic Frameworks

*Sandro Wieser, Tomas Kamencek, Johannes P. Dürholt, Rochus Schmid, Natalia Bedoya-Martínez\*, Egbert Zojer\**

S. Wieser, T. Kamencek, Prof. E. Zojer
Institute of Solid State Physics, Graz University of Technology, NAWI Graz, Petersgasse 16, 8010 Graz, Austria
E-mail: egbert.zojer@tugraz.at

T. Kamencek
Institute of Physical and Theoretical Chemistry, Graz University of Technology, NAWI Graz, Stremayrgasse 9, 8010 Graz, Austria

Dr. J. P. Dürholt, Prof. R. Schmid
Computational Materials Chemistry Group, Faculty of Chemistry and Biochemistry, Ruhr-University Bochum, Universitätsstraße 150, 44801 Bochum, Germany

Dr. N. Bedoya-Martínez
Materials Center Leoben, Roseggerstraße 12, 8700 Leoben, Austria
E-mail: OlgaNatalia.Bedoya-Martinez@mcl.at

Keywords: metal-organic framework (MOF), thermal conductivity, thermal interface resistance, non-equilibrium molecular dynamics (NEMD)

Controlling the transport of thermal energy is key to most of applications of metal-organic frameworks. Analyzing the evolution of the effective local temperature, the interfaces between the metal nodes and the organic linkers are identified as the primary bottlenecks for heat conduction. Consequently, changing the bonding strength at that node-linker interface and the mass of the metal atoms can be exploited to tune the thermal conductivity. This insight is generated employing molecular dynamics simulations in conjunction with advanced, ab initio parametrized force fields. The focus of the present study is on MOF-5 as a prototypical example of an isoreticular MOF. Still, the key findings prevail for different node structures and node-linker bonding chemistries. The presented results lay the foundation for developing detailed structure-to-property relationships for thermal transport in MOFs with the goal of devising strategies for the application-specific optimization of heat conduction.

The structure of a Metal-Organic Framework (MOF) is characterized by an open framework of metal ions interconnected by organic linkers. Its porous structure is particularly interesting for various applications including catalysis,[1] and the capture, controlled release,[2] storage,[3] and separation of gases.[4] During gas capture and release processes heat is either released or consumed. Similar considerations apply to reactions catalyzed by MOFs. For such applications, MOFs with maximized thermal conductivities are highly desirable, to either maintain isothermal conditions or to realize temperatures necessary for sufficiently high reaction rates.[5] The opposite applies to MOFs used in thermoelectric devices,[6] for which particularly low thermal conductivities are needed. These examples show that the suitability of a MOF for a specific application crucially depends on its ability to transport heat.

The sheer number of possible organic linkers, metal nodes, topologies, and framework architectures implies that there should be MOFs with hugely varying thermal transport properties. To exploit this variability, it is imperative to make the right choice from a pool of tens of thousands of MOF structures that have been realized so far (and from potentially millions that could be made, if required).[7] This calls for the development of detailed structure-to-property relationships for thermal transport, linking MOF structures and their thermal transport properties. Currently, such an in-depth understanding is still elusive. To date, most of the experimental and theoretical work on thermal transport in MOFs has focused on MOF-5 (shown in **Figure 1**a), also known as IRMOF-1. It constitutes an isoreticular cubic framework consisting of $ZnO_4$ clusters connected by 1,4-benzenedicarboxylate (BDC) linkers,[8] with a comparably low thermal conductivity of 0.32 W/m K at 292 K measured on single crystal samples in nitrogen atmosphere.[9] Due to its simple structure and cubic symmetry, MOF-5 in the present contribution serves as a prototypical system for studying the specific role of nodes and linkers for thermal transport. To show the broader validity of our results, we also show selected data on MOF-508 (displayed in Figure 1b). MOF-508 is an example for a paddle wheel based, pillared layer MOFs. It differs from MOF-5 in symmetry, and more importantly, in the structure and shape of the nodes. These consist of $Zn_2O_8$ paddle wheels, connected in the x-y-plane by BDC linkers and in z-direction by bipyridines. This yields a highly anisotropic bonding chemistry between nodes and linkers. MOF-508 is also of practical relevance for the separation of alkanes,[10] and as a configurable base system for high $CO_2$ uptake,[11] while MOF-5 is primarily interesting for $CO_2$ and $H_2$ gas-storage.[12] At this point it should be stressed that our focus is on pristine MOF structures to elucidate their fundamental properties. Therefore, we will not consider the impact of gas adsorption and guest infiltration, which is known to affect heat transport properties.[13] We will also disregard interpenetrated systems, even though they often occur for MOF-508-type materials.[14]

Considering the modular structure of a MOF, it is crucial to first understand, which part of the framework most strongly impacts thermal transport. Therefore, we will provide a spatially resolved view of the thermal resistivity within the considered MOF structures. To achieve that, we impose a temperature gradient over a supercell consisting of a series of MOF unit cells, as commonly done in non-equilibrium molecular dynamics (NEMD) simulations to obtain thermal conductivities (see Figure 1c).[15] Subsequently, we analyze the progression of the effective local temperature, $\overline{T}'$, which is defined as $2/(3k_B)$ times the average kinetic energy in small regions encompassing equivalent atoms in planes perpendicular to the direction of heat flux. This averaging is carried out over at least 20 million time steps (10 ns).

Considering the sheer size of the required supercells (comprising up to tens of thousands of atoms) and the number of distinct systems discussed here, the NEMD simulations were performed employing classical force fields (in conjunction with the LAMMPS code).[16] To maximize the accuracy of the obtained results, we refrained from using a generic force field, like for example UFF,[17] but used the second-generation MOF-FF,[18] which has been systematically parameterized for each of the considered systems against high-level, first-principle DFT data. The latter have been calculated with VASP for the full, 3D periodic structure of the MOFs.[19] MOF-FF was previously used to successfully predict other phonon dependent properties, like negative thermal expansion (NTE) effects of MOFs.[20] Besides, we recently showed that MOF-FF is capable of describing the phonon properties of molecular crystals (in particular naphthalene) with an accuracy close to its reference DFT method, outperforming not only other force fields but also DFTB methods.[21] To obtain thermal conductivities, finite size effects were accounted for by extrapolating from data for supercells of increasing size.[15a] Conversely, whenever reporting effective local temperatures, these are reported for a specific supercell size. Further details on the force field parametrization and benchmarking, on the NEMD simulations, and on the extrapolation to infinite-size samples can be found in the Supporting Information.

Employing the procedure described above, we obtained a thermal conductivity, $\kappa$, of (0.42 ± 0.04) W(mK)$^{-1}$ at 300 K for MOF-5 with Zinc-based nodes when using the MOF-FF. As detailed in the Supporting Information, a similar value of (0.38 ± 0.03) W(mK)$^{-1}$ is calculated employing a recently published (numerically much more demanding) neural network potential specifically developed for MOF-5,[22] corroborating the accuracy of the MOF-FF data. Both values are somewhat larger than the experimental result of 0.32 W(mK)$^{-1}$ at 292 K.[9] This is not surprising, considering that in the simulations a perfect, defect-free structure is assumed, while the defects present in actual samples typically decrease $\kappa$.[23] Employing the complementary Green Kubo approach, Huang et al. obtained a similar value for κ (0.31 W(mK)$^{-1}$).[24] For MOF-508 (Zn), the thermal conductivity becomes anisotropic due to the reduced symmetry. For heat flow parallel to the BDC and bipyridine pillars the values of κ amount to 0.59 and 0.49 W(mK)$^{-1}$, respectively. Here, a comparison to experiments to the best of our knowledge is not possible due to a lack of heat-transport related data.

Considering the good overall agreement of our simulations with the experimental data for MOF-5 (Zn), we next turn to analyze the positional dependence of the effective local temperature (see Figure 1d). The data reveal that $\bar{T}'$ does not change continuously with position. Instead, there is a plateau with a very small temperature gradient in the region of the organic linker. This implies a high thermal conductivity in that part of the MOF. The temperature gradient in the node is higher than in the linker, but the by far most pronounced drop in $\bar{T}'$ occurs at the interface between the node and the linker. This suggests that the main heat transport bottleneck in MOF-5 is the interface between organic and inorganic building units. The same trend is found for MOF-508 despite changes in the structure of the node and the bonding chemistry between the node and the linker (see Figures 1e and f). Such interfacial thermal resistances, also referred to as Kapitza resistances, are common at heterogeneous interfaces between bulk materials,[25] and have also been proposed for monolayer junctions between metallic electrodes.[26] Notably, for carbon nanotube networks and stacks of semiconducting or dielectric heterolayers, interfacial resistances have also been identified via jumps in the local temperature between individual tubes,[27] or the different semiconducting and dielectric materials.[28] To the best of our knowledge, the present case is, however, the first observation of such an effect within a neat bulk material. Moreover, the interfaces causing the resistance have neither been created by deliberately engineering semiconductor superlattices,[29] nor are they the consequence of a random arrangement of building blocks (like in the above-mentioned sheets of the carbon nanotubes). Instead, they originate from a 3D self-assembly process that generates the interfaces in a bottom-up fashion, when combining the individual starting materials during MOF growth.

This raises the question, how the node-linker interface resistance could be tuned. Considering that it originates from the differences in phonon properties of linkers and nodes, a straightforward approach would be to change the metal atoms (and their masses) in the nodes.

The impact of different metals (Mg, Ca and Zn) on the profiles of the effective local temperature in MOF-5 is shown in **Figure 2**a. In contrast to Figure 1, we here plot the temperature profiles of consecutive units on top of each other (aligning them to the temperatures of the center of the respective nodes). Figure 2a also contains the corresponding linear regression fits performed to obtain the temperature gradients in the respective segments. The plot shows, that in all three systems the situation is qualitatively similar, i.e., the temperature profiles are rather flat in the regions of the linker, steeper across the nodes and there is a jump at the interface. This is also the case in Figure 2b for MOF-508, although there the contribution of the node is difficult to identify for thermal transport along the direction of the BDC linkers. This is a consequence of the negligible spatial extent of the nodes in that direction. As the temperature difference between hot and cold thermostats is the same for all simulations, one has to go beyond the temperature profiles to appreciate the differences between the materials. Thus, Figure 2a and Figure 2b also list the values of the calculated heat flow per area, $j$. One sees that $j$ increases somewhat from MOF-5 (Ca) to MOF-5 (Zn) and then grows significantly for MOF-5 (Mg). Moreover, for MOF-508 in the direction of the BDC linker, $j$ is considerably higher than for bipyridine.

From the heat flow per area and the drop in the effective local temperature between sample positions $z_a$ and $z_b$ (derived from a linear fit), one can calculate the local thermal resistance, $R_{ab}^{th}$, between these two positions. In the following, we will report $R_{ab}^{th}A$ rather than $R_{ab}^{th}$ to obtain values that independent of the actual cross-section of the sample, $A$, and to be also consistent with the definition of the Kapitza interface resistance,[30] $R_{Kapitza} = R_{interface}^{th} \cdot A$. The interface resistance is obtained from the drop in temperature at the interface between the nodes and the linkers. Then, the resistance of a "thermal repeat unit", $R_{unit}^{th}$, consisting of one node with a thermal resistance $R_{node}^{th}$ and a linker with $R_{linker}^{th}$ is given by:

$$R_{unit}^{th} A = \frac{\Delta z_{unit}}{\kappa} = \left(R_{node}^{th} + R_{linker}^{th} + 2 \cdot R_{interface}^{th}\right) A \qquad (1)$$

Here, $\Delta z_{unit}$ is the length of the thermal repeat unit. Rigorous definitions of all quantities from Equation (1) and details on how they are calculated can be found in the Supporting Information.

The resulting values of the local thermal resistances are plotted in Figure 2c. These data confirm the notion that the lowest contribution to the thermal resistance always stems from the linker and that the interfaces are, indeed, the main bottlenecks for thermal transport. In the MOF-5 based systems, the thermal resistance of the node can be either similar to that of the linkers (in the case of MOF-5 (Mg)) or nearly twice as large (for MOF-5(Ca)). This ratio substantially increases in MOF-508 (Zn) along the direction of the bipyridine units, as their thermal resistance is particularly low. This is consistent with the largely vanishing temperature gradient in that direction in the region of the linker (see Figure 2b) and could be an interesting starting point for maximizing the thermal conductivity. For the specific case of MOF-508 (Zn), its particularly small value is, however, compensated by a large interface resistance.

As far as the nodes and interfaces are concerned, one finds significantly different values already in the MOF-5 based systems. The values of $R_{node}^{th}$ and $2 \cdot R_{interface}^{th}$ more than double from MOF-5 (Mg) to MOF-5 (Ca). Consequently, MOF-5 (Ca) displays the largest value of $R_{unit}^{th}$. Consistent with this observation, Han et al. found by scaling the mass of the entire nodes (leaving all other force-field parameters the same) that decreasing the mass considerably increases the thermal conductivity.[31] This has been attributed to the reduction of the mass mismatch between node and linker, which is expected to reduce phonon scattering at the interface due to an increased overlap in the phonon density of states of the two subsystems.[32] This argument would explain the change from Mg to Ca, but is at variance with the observation that the even heavier Zn-based nodes result in a reduced interface as well as unit resistance compared to Ca. For MOF-508 (Zn) in BDC direction, one even observes the smallest unit resistance of all considered systems, which in conjunction with the shortest length of the thermal repeat unit in that MOF results in essentially the same thermal conductivities for MOF-508 (Zn) in BDC direction and MOF-5 (Mg) (see **Figure 3**a).

The above arguments do, however, not explain the observed evolution of the thermal resistance with the type of metal in the MOF-5 based systems. To disentangle the impact of the mass and possible variations in the bonding interaction, we performed NEMD simulations for all MOF-5 based systems, varying the metal mass, but leaving all force-field parameters at the values for the Mg, Ca, or Zn based systems. The obtained trends in thermal conductivities corrected for finite-size effects, are shown in the left panel of Figure 3a. Indeed, reducing the mass leads to an increase in thermal conductivity, while increasing it beyond 70 u has only a minor impact. As shown in Figure 3b, the drop in thermal resistance at low masses is primarily a consequence of a pronounced decrease of the interface resistance around 30-40 u. Interestingly, this is equal to about twice the mass of a bonded oxygen atom, which is reminiscent of observations for interfaces between simple model systems, where a sharp change in thermal conductance is observed in the same relative mass range.[32b]

For MOF-5 (Ca) derived systems, despite trends consistent with the other materials, the absolute values of the different thermal resistances increase by a factor of ~2 larger. According to Figure 3b, this is again primarily caused by a distinct increase of $R^{th}_{interface}$ for the MOF-5 (Ca) force field, which suggests a different bonding chemistry than for Mg and Zn. Indeed, in MOF-5 (Ca) the metal-oxygen bond length amounts to 2.25 Å, which is distinctly larger than in the 1.97 Å obtained for MOF-5 (Zn) and the 1.96 Å for MOF-5 (Mg). Also, the associated reference force constants are significantly smaller (see the Supporting Information). To show that a comparably weak Ca-O bond is the reason for the particularly low thermal conductivity in MOF-5 (Ca), we explored the impact of rescaling the corresponding parameter in the respective force field. For an increased force constant, this indeed yields the expected decrease of the interface resistance, as shown explicitly in the Supporting Information.

In summary, we find that the bottlenecks to thermal transport in metal-organic frameworks are the interface between the nodes and the linkers, which give rise to a Kapitza resistance dominating thermal transport. This is manifested in a massive drop of the effective local temperature at these interfaces, which prevails when changing the structure of the nodes, the bonding chemistry between nodes and linkers, as well as the masses of the metal atoms forming the nodes. The latter is identified as a handle for tuning the interface resistance, where a reduction in the mass mismatch between node and linker, results in a higher thermal conductivity associated with decreased phonon scattering. Changing the bonding strength between node and linker is identified as another handle for tuning the Kapitza resistance. Exploiting interface resistances for tuning heat transport is a strategy also pursued in semiconductor superlattices, where phonon wave interference can lead to an abnormal thermal conductivity behavior.[33] A distinct advantage of MOFs in this context is that they contain periodic assemblies of interfaces that do not need to be realized in a top-down approach but are generated "naturally" during the self-assembly of the material.

**Supporting Information**

Supporting Information is available from the Wiley Online Library or from the author.

**Acknowledgements**

E.Z., T.K. and S.W. acknowledge the Graz, University of Technology for financial support through the Lead Project (LP-3). T.K. acknowledges additional funding as recipient of a DOC Fellowship of the Austrian Academy of Sciences at the Institute of Solid State Physics. N. B.-M. acknowledges the financial support under the scope of the COMET program within the K2 Center "Integrated Computational Material, Process and Product Engineering (IC-MPPE)" (Project No 859480). This program is supported by the Austrian Federal Ministries for Climate Action, Environment, Energy, Mobility, Innovation and Technology (BMK) and for Digital and Economic Affairs (BMDW), represented by the Austrian research funding association (FFG), and the federal states of Styria, Upper Austria and Tyrol. J.P.D. and R.S. acknowledge financial support from the Deutsche Forschungsgemeinschaft (DFG, Grants: SCHM 1389/10-1 within FOR 2433 and SCHM 1389/8-1). The Vienna Scientific Cluster (VSC3) is acknowledged for providing the computational resources required for this project. We would like to thank also Jörg Behler and Marco Eckhoff for providing the neural network potential used in this work.

Received: ((will be filled in by the editorial staff))
Revised: ((will be filled in by the editorial staff))
Published online: ((will be filled in by the editorial staff))

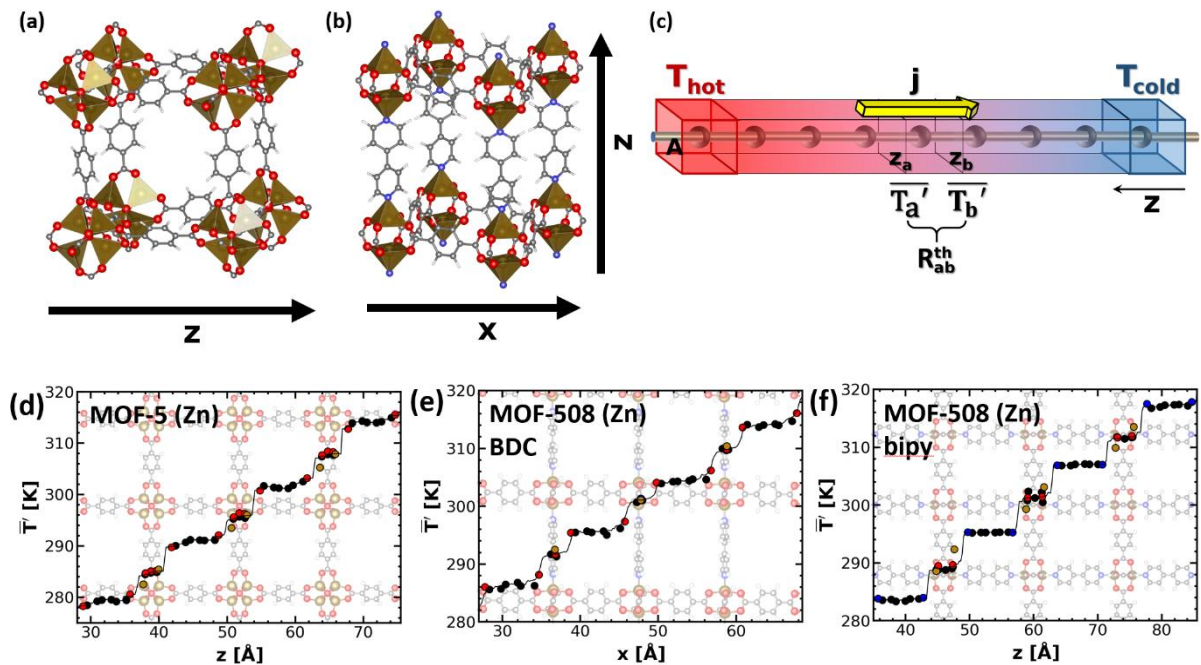

**Figure 1**. (a) and (b) show the atomistic structure of MOF-5 and MOF-508 around one of their pores. [Zn: gold; O: red; N: blue; C: grey; H: white]; (c) contains a schematic representation of a NEMD simulation. Thermostats introduce regions with increased temperature $T_{hot}$ and reduced temperature $T_{cold}$, which results in a heat flux $j$ across the system. In a certain region

$z_a$-$z_b$ the difference in local temperatures $\overline{T'_a}$ and $\overline{T'_b}$ can be used to obtain the local thermal resistance $R_{ab}^{th}$. (d-f) show sections of the effective temperature profile obtained in a NEMD simulation for cells with a length of 16 nodes. In (d), the profile of MOF-5 (Zn) is shown, while (e) and (f) visualize the temperature profile of MOF-508 in directions of benzenedicarboxylate (x) and bipyridine (z) linkers, respectively. The symbols represent the effective local temperature obtained by averaging the kinetic energies of equivalent atoms (omitting hydrogens in the plot for the sake of visual clarity). The solid lines represent the moving average of the effective local temperature over a z-range of 3 Å.

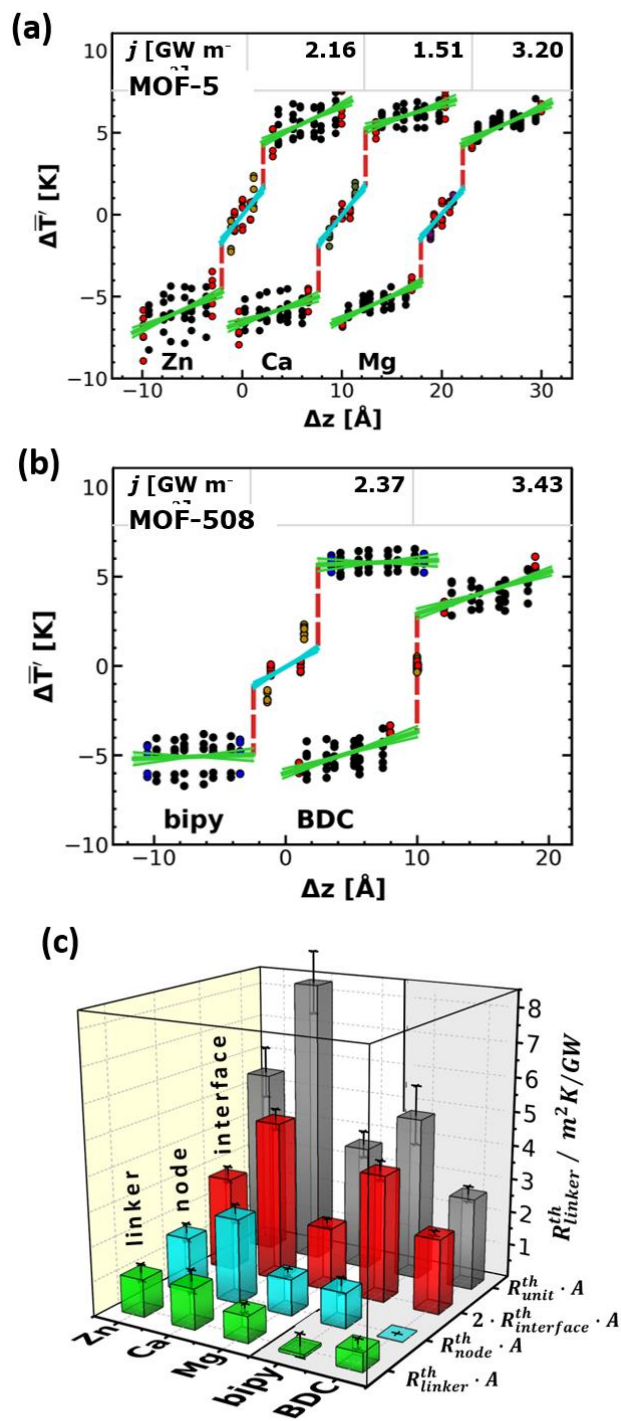

**Figure 2**. Overlaid profiles of the effective local temperatures of neighboring units of MOF-5 (a) and MOF-508 (b). Temperatures are reported relative to the value in the center of the linker. For MOF-5 (MOF-508), supercells consisting of 8×1×1(16×4×4) unit cells have been considered with the largest extent of the supercells along the direction of heat flow. The larger

nominal size of the MOF-508 supercells is due to its smaller unit cell and a slower convergence regarding cell thickness (see the Supporting Information). Linear fits to the temperature evolutions have been performed across the linkers (green) and nodes (blue). The remaining temperature differences are represented as steps at the linker-node boundaries (red). The linkers perpendicular to the heat flux (apart from their terminal O atoms) have been disregarded in the analysis (for reasons explained in the Supporting Information). The heat fluxes $j$ are listed for each system. (c) compares the key contributions to the thermal resistances for MOF-5 variants and for MOF-508 (for a definition of the quantities see the main text). The error bars result from a 99 % confidence interval of the errors of the linear fits as illustrated by the spread of the linear fits in panels (a) and (b).

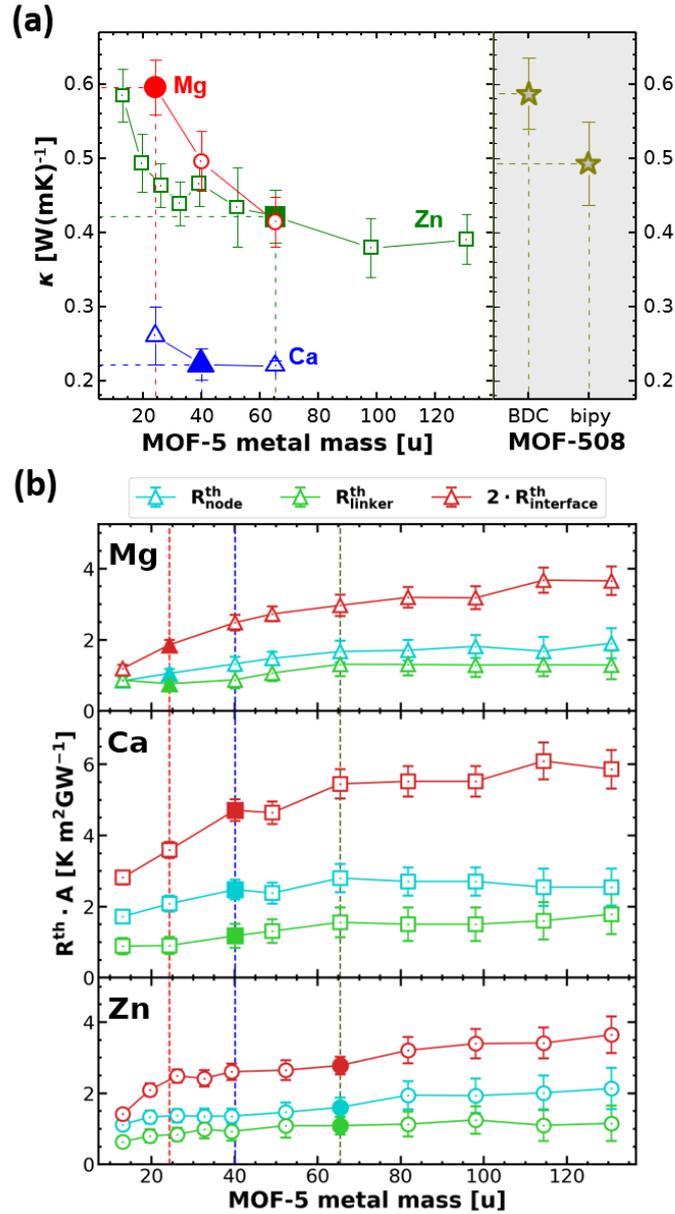

**Figure 3**. (a): Thermal conductivities calculated for MOF-5 (Mg) (circles), MOF-5 (Ca) (triangles), and MOF-5 (Zn) (squares) upon scaling the mass of the metal atoms in the nodes in the NEMD simulations. The error bars result from a 95 % confidence interval of the standard deviation from the linear fit to the infinite size limit. For the sake of comparison, the thermal conductivities of MOF-508 in direction of the BDC and the bipyridine linkers are also shown. (b): Thermal resistance contributions of node, linker and interfaces for MOF-5 variants with scaled metal masses for an 8×1×1 supercell. Filled symbol in (a) and (b) denote the actual mass of the metal atom for which the respective force field has been parametrized.

TOC

The interface between the inorganic node and the organic linker is identified as the primary bottleneck for heat transport in metal-organic frameworks employing non-equilibrium molecular dynamics simulations. Its extent can be tuned by the choice of metal and the node-linker bonding strength, providing the tools for tuning thermal conductivities to the needs of specific applications.

**Thermal Transport in MOFs**

S. Wieser, T. Kamencek, J. P. Dürholt, R. Schmid, N. Bedoya-Martínez*, E. Zojer*

**Identifying the Bottleneck for Heat Transport in Metal-Organic Frameworks**

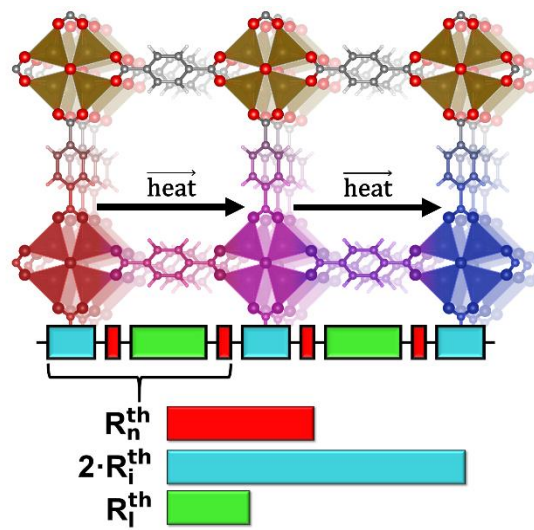

Supporting Information

**Identifying the Bottleneck for Heat Transport in Metal-Organic Frameworks**

*Sandro Wieser, Tomas Kamencek, Johannes P. Dürholt, Rochus Schmid, Natalia Bedoya-Martínez\*, Egbert Zojer\**

# S1. Obtaining DFT reference data

The force fields have been parametrized based on reference data from density functional theory (DFT) calculations. These were carried with the Vienna Ab-Initio Simulation Package *VASP* (version 5.4.4)[1–4] employing the following standard pseudopotentials[5,6] for the PBE functional[7] within the projector-augmented wave method. Note that for Ca, the sv pseudopotential was used, which treats the 3s and 3p electrons as valence states. This was done to be able to use a larger plane wave energy cutoff required for convergence. The specific versions of the pseudopotentials used for the individual elements can be seen in Table S1.

**Table S1.** PAW pseudopotential used for the individual elements for the reference calculations.

| Element | PAW Pseudopotential Title |
| --- | --- |
| H | PAW_PBE H 15Jun2001 |
| C | PAW_PBE C 08Apr2002 |
| N | PAW_PBE N 08Apr2002 |
| O | PAW_PBE O 08Apr2002 |
| Mg | PAW_PBE Mg 13Apr2007 |
| Ca | PAW_PBE Ca_sv 06Sep2000 |
| Zn | PAW_PBE Zn 06Sep2000 |

The occupation of electronic states was described with a Gaussian smearing with a distribution width of $\sigma = 0.05$ eV. The sampling of the reciprocal space and the plane wave energy cutoff were adjusted for each system to converge the total energy below 1 meV per atom. We have employed this approach to obtain highly reliable and converged vibrational properties in the past.[8,9] The system-specific settings can be found in Table S2.

**Table S2.** System-specific cutoff and mesh sampling settings used for the DFT reference calculations

| System | $k$-mesh sampling (Γ-centered) | Plane Wave Energy Cutoff [eV] |
| --- | --- | --- |
| MOF-5(Zn) | 2×2×2 (→ 3 irreducible $k$-points) | 800 |
| MOF-5(Mg) | 2×2×2 (→ 3 irreducible $k$-points) | 800 |
| MOF-5(Ca) | 2×2×2 (→ 3 irreducible $k$-points) | 800 |
| MOF-508 | 3×3×2 (→ 9 irreducible $k$-points) | 900 |

For all systems, an SCF energy convergence criterion of $10^{-8}$ eV and the global precision parameter *Accurate* (for details see the *VASP* manual[10]) were used (keyword `PREC`). Moreover, projection operators were evaluated in reciprocal space (*VASP* keyword `LREAL` set to `False`), which we found to be crucial to obtain reliable frequencies. Grimme's D3 dispersion correction with Becke-Johnson damping (D3-BJ)[11,12] was employed to treat van der Waals interactions in the systems. As was pointed out by some of us in a detailed study about crystalline naphthalene,[8] the PBE functional in combination with the D3-BJ correction results in highly accurate lattice parameters and phonon properties in all frequency ranges.

For fitting atomic charges with the REPEAT[13] method for the force fields (FFs), the local potential (*LOCPOT*) was generated with *VASP*. These local potentials were explicitly chosen to include only electrostatic contributions (ionic and Hartree), to omit exchange and correlation contributions (by setting the variable `LVHAR` to `True`).

The PHONOPY[14] code was used to calculate the interatomic force constants and vibrational Γ-phonon modes using primitive unit cells and the default displacement distance of 0.01 Å, which we have usually found suitable for MOFs,[9] and organic semiconductors.[8] The harmonic force constants were symmetrized with PHONOPY's internal routines to account for the acoustic sum rules.

## S2. System geometries

The DFT reference calculations and the fits of the force fields were carried out as described in section S3 using the primitive cell of MOF-5 for Mg, Ca, and Zn. The system is equivalent to a fourth of the conventional face-centered-cubic (fcc) cell and contains 106 atoms in two nodes and 6 linkers. It is visualized in Figure S1a (all atomistic structures used in this work were generated using the VESTA software).[15] The conventional cubic unit cell, containing 424 atoms, is used as a basis for all the molecular dynamics simulations and includes 8 nodes and 24 linkers and is shown in Figure S1b. This large unit cell is the result of two different alignments of the phenyl rings in two consecutive linkers. This results in neighboring nodes and linkers to be geometrically inequivalent.

In contrast to MOF-5, for MOF-508, the cells used in the DFT reference calculation and as basic repeat unit in the molecular dynamics simulations are the same. They show an orthorhombic symmetry (space group C222) and are visualized in Figure S1c. At equilibrium, the torsion angle in the bipyridine linker amounts to 36.87°, and the phenylene in the BDC (1,4-benzenedicarboxylate) linkers are inclined by 2.6° relative to the plane of metal atoms. This small deviation from full planarity, as is the case for MOF-5, is most likely due to the proximity and torsion of the bipyridine pillar linkers.

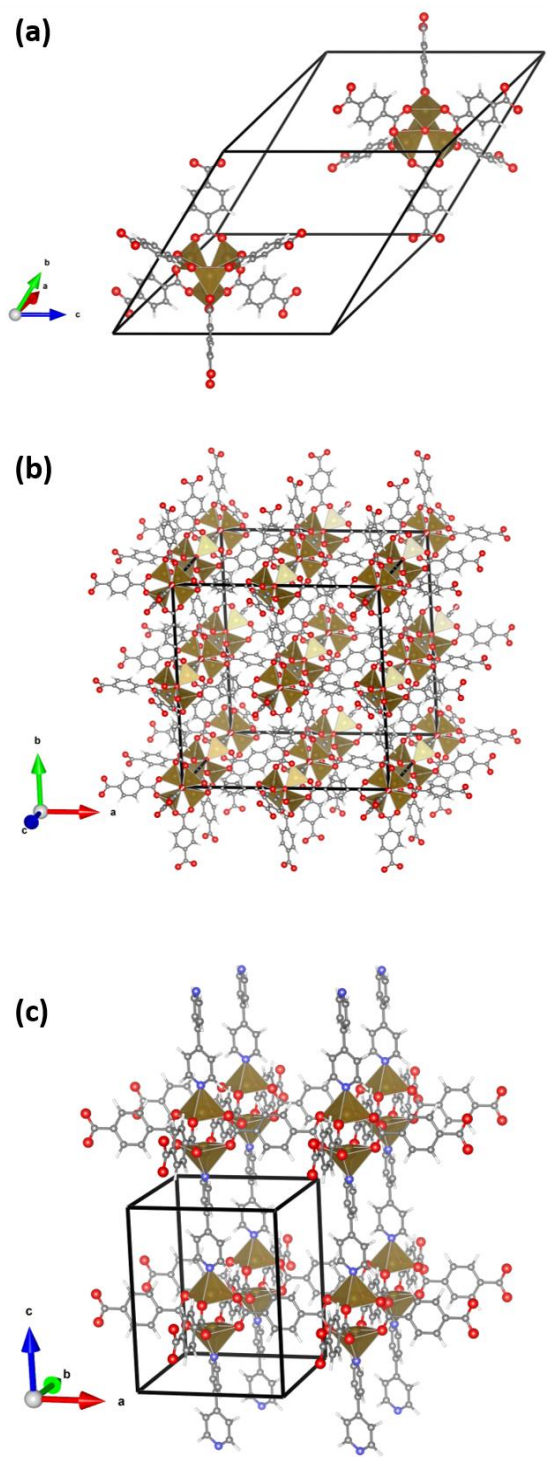

**Figure S1.** Structures used throughout this work. (a) Primitive unit cell of MOF-5, incomplete linkers have been expanded for clarity. The cell only contains two full nodes and six linkers. (b) Conventional fcc unit cell of MOF-5, incomplete node-linker clusters at the boundary have been expanded for clarity. The cell encompasses eight full nodes and 24 linkers. (c) Unit cell of MOF-508; the indicated area represents one primitive cell. The cell includes one full node, two BDC linkers and one bipyridine linker. grey: carbon, white: hydrogen, red: oxygen, blue: nitrogen, gold: metal

## S3. Fitting of the force fields

The force fields were fitted with FFgen[16] including all terms also contained in the original MOF-FF potential[17] and an additional torsional cross term, which is considered in several class-2 force fields:

$$E_{bb13} = N(r_{ij} - r_1)(r_{kl} - r_3) \qquad (S1)$$

This term describes the interaction between two next-nearest bonds $r_1$ and $r_3$ of connected atoms as visualized in Figure S2. Van der Waals interactions were not directly fitted but taken from the MM3 force field.[18] Atomic charges were obtained from the electrostatic potential of the reference calculations DFT calculations (see above) using the REPEAT method.[13]

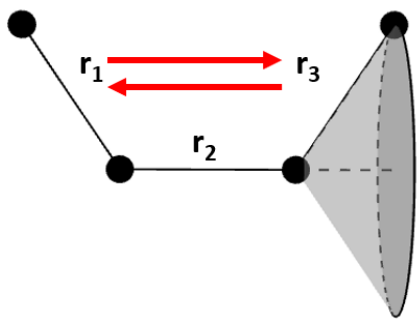

**Figure S2.** Visualization of the bb13 interaction, including 4 connected atoms. The additional term describes the interaction of the distance $r_1$ with the distance $r_3$.

The functional form of the FFs for the different MOF-5 variants (containing different metals in the nodes) is identical. All fits were performed for at least 5000 CMA-ES (covariance matrix adaptation - evolution strategy) generations with a population size of 48 each to achieve proper convergence. Section S4 contains a detailed analysis of the force field quality compared to various available reference quantities.

The actual force field parameters obtained this way can be found in a GitHub repository accompanying this work at https://github.com/sandrowieser/mof_thermal_bottleneck_si.

## S4. Benchmarking the quality of the parametrized force fields

This section aims at providing an overview of how well the individual force fields agree with the reference data and how well they can be expected to describe phonon-related properties, and therefore how suited they are to estimate thermal conductivities. For this, we look at phonon properties obtained via lattice dynamics simulations carried out with the help of the phonopy package.[14] The required single-point calculations and cell relaxations to obtain the data for the force fields were performed with the LAMMPS code.[19]

### S4.1 Quality of geometrical parameters and forces

The force fields are fitted to DFT geometries and force constants. In Figure S3 a comparison of the geometrical parameters can be seen for MOF-5 (Zn). The agreement is almost perfect and looks identical for the other MOF-5 variants containing Ca and Mg. In Figure S4, the agreement of the force constants in redundant internal coordinates (where the Cartesian Hessian matrix is converted to additionally describe interactions between bonds, angles and torsions)[20] can be seen. For MOF-508, the agreement is worse, which we attribute to the additional complexity introduced by the slightly inclined linkers (as described in section S2) and to the fact that some of the cross-terms were not included in the fitting process to simplify the parametrization. The omitted terms describe comparably insignificant interactions in the node (these specifically occur within interactions that include at least three metal or oxygen atoms, to see precisely which interactions were included, refer to the full list of parameters contained in the accompanying GitHub repository), which would lead to a significant increase in the number of parameters to fit. This should not change the accuracy of the obtained heat transport properties, as these terms mostly lightly affect higher frequency localized modes that contribute little to the thermal conductivity. The root-mean-square errors (RMSEs) for the above-described parameters are listed in Table S3 for all MOF parametrizations. This includes the agreement for the low-frequency modes up to 300 cm$^{-1}$, which are the low-frequency modes typically most important for heat transport.[21] The data in Table S3 confirm the excellent agreement for MOF-FF calculated properties of MOF-5 compared to the DFT reference data. For MOF-508, the agreement is significantly worse, and the RMSE of vibrational frequencies amounts to 31.9 cm$^{-1}$. Although this is inferior to the MOF-5 based systems, the agreement is still better than what has been reported for standard transferable classical force fields, like, for example, COMPASS.[22]

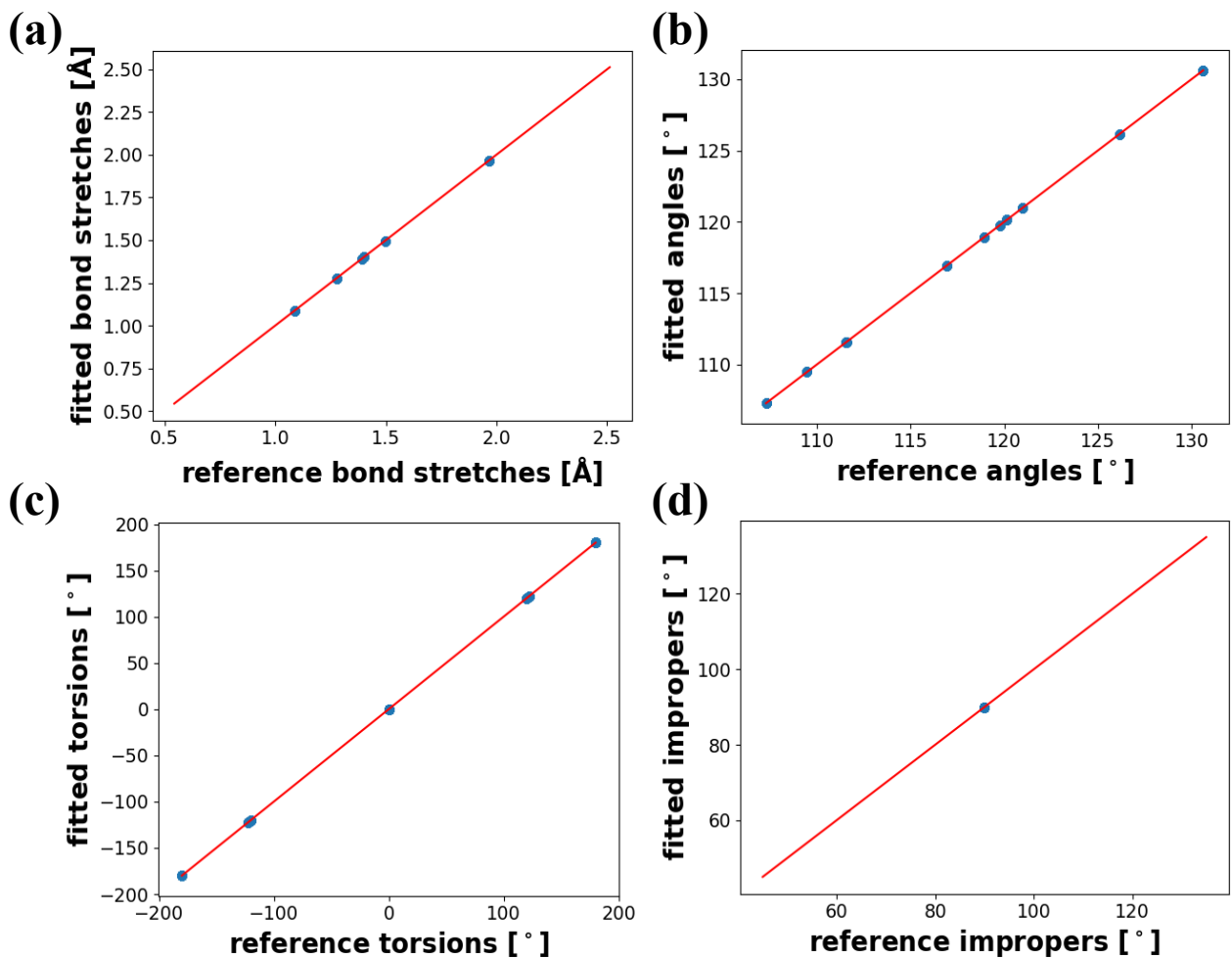

**Figure S3.** Comparison of the fitted MOF-5 (Zn) force field to DFT reference data for geometrical properties: (a) bond lengths between two atoms, (b) in-plane bending angles between three bonded atoms, (c) proper torsion angles describing the bending out of the planar configuration of four linearly bonded atoms (see Figure S2 for visualization) and (d) improper torsion angles describing the out of plane bending of the central atom bonded to three neighboring atoms.

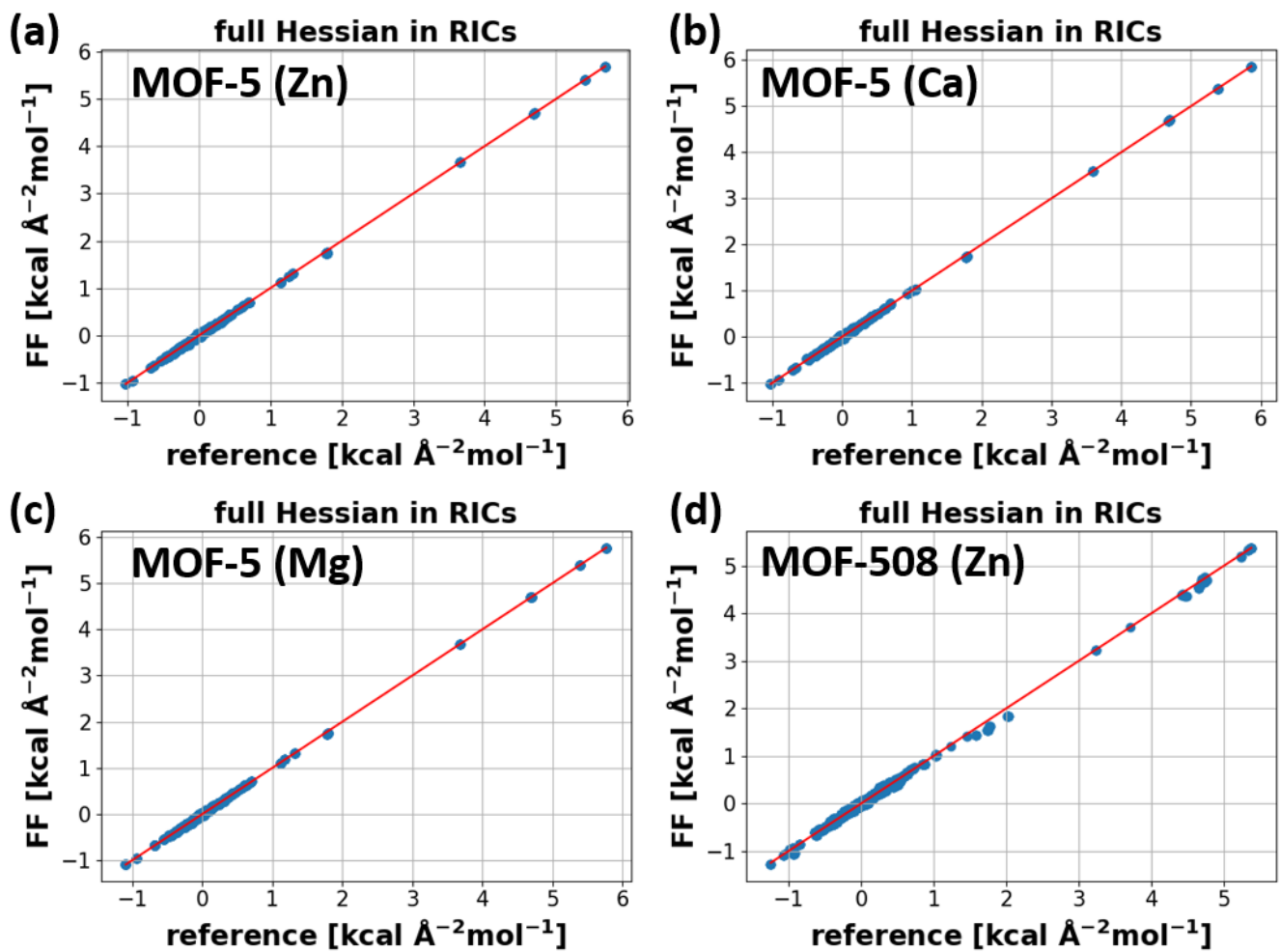

**Figure S4.** Comparison of the fitted FF to DFT reference Hessians for MOF-5 (Zn) (a), MOF-5 (Ca) (b), MOF-5 (Mg) (c) and MOF-508 (Zn) (d) in redundant internal coordinates (RICs), which describe the interactions between individual bonds, angles and torsions.

**Table S3.** Root mean square errors of properties in the force field compared to the DFT reference.

|  | MOF-5 (Mg) | MOF-5 (Ca) | MOF-5 (Zn) | MOF-508 (Zn) |
|---|---|---|---|---|
| bond lengths [Å] | 0.00009 | 0.00009 | 0.0001 | 0.004 |
| angles [°] | 0.005 | 0.003 | 0.006 | 0.14 |
| proper torsions [°] | 0.008 | 0.002 | 0.005 | 0.21 |
| improper torsions [°] | 0.004 | 0.000002 | 0.001 | 0.20 |
| hessian [kcal Å$^{-2}$ mol$^{-1}$] | 0.002 | 0.002 | 0.003 | 0.004 |
| frequencies [cm$^{-1}$] | 13.6 | 13.7 | 13.9 | 31.9 |
| frequencies (≤300 cm$^{-1}$) [cm$^{-1}$] | 4.4 | 5.9 | 7.1 | 16.1 |

## S4.2 Benchmarking of vibrational properties

As a first step, we compare the Γ point vibrational frequencies calculated with our MOF-FF parametrizations with the corresponding DFT reference data. The results are contained in Figure S5. Since the frequencies are not necessarily in the same order for the force field and DFT simulations, it is necessary to sort the vibrations according to the associated displacement patterns. When such a sorting was not performed, it would frequently occur that a mode from the reference is compared to a completely different FF-calculated mode, which just happens to be at a similar frequency. This would lead to a systematic underestimation of the error. To avoid this, all eigenvectors of the MOF-FF and DFT calculated modes were compared via dot products, and the best agreement was determined with the help of a minimization problem in the form of the Hungarian method for the assignment problem.[23] This leads to RMSE values, as indicated in Table S3. In Figure S6, histograms of the eigenvector agreement are visualized. Here, it can be seen that for MOF-5, many modes agree very well with the reference. It also becomes apparent that some of the modes are not described that well. Many of the latter are, however, fairly localized modes (like, for example, rotations of the bipyridine linkers around their axis). The degree of localization can be quantified via participation ratios, where modes with a low participation ratio ave been shown to have a low group velocity for typical MOFs[9]. This suggests a reduced relevance of these modes for heat transport.

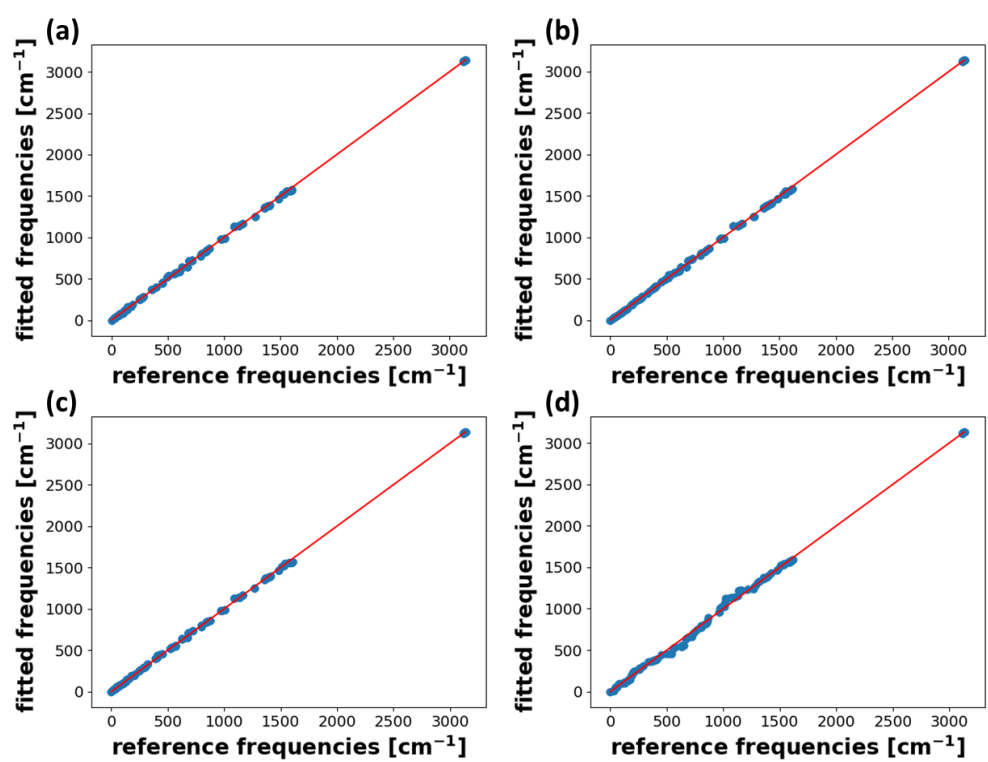

**Figure S5.** Agreement of vibrational frequencies calculated with the FF at the Γ point compared to the DFT reference. (a) MOF-5 (Zn) (b) MOF-5 (Mg) (c) MOF-5 (Ca) (d) MOF-508

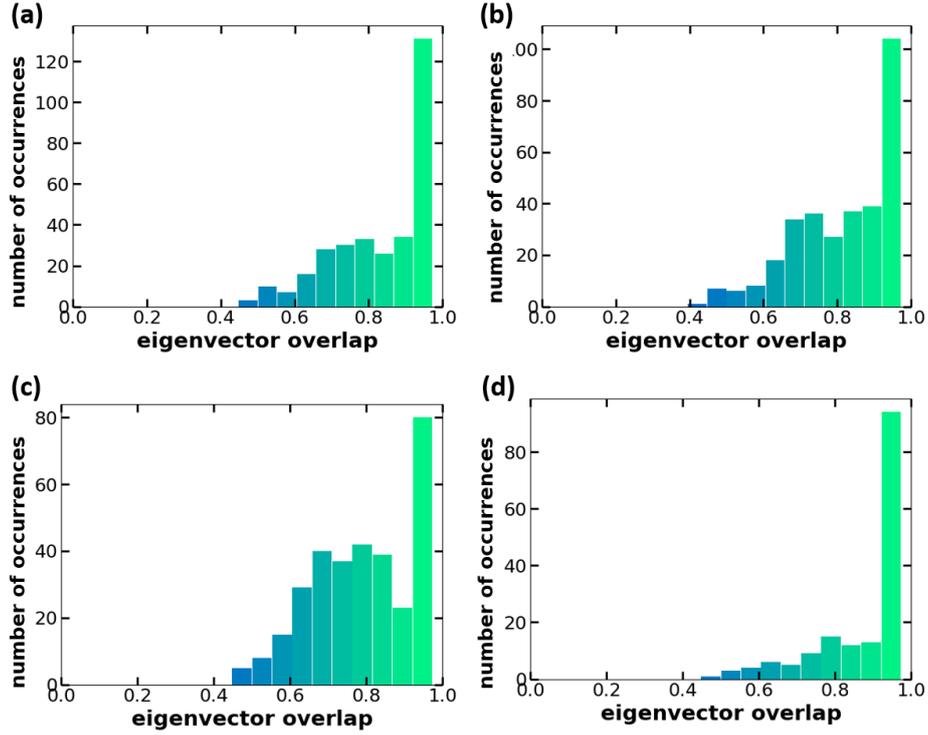

**Figure S6.** Histograms of the dot products (eigenvector overlaps) of the eigenvectors of associated Γ point vibrations calculated with DFT and with the fitted force field. Here, eigenvectors calculated with the two approaches, that maximize the dot products, are associated to each other. (a) MOF-5 (Zn) (b) MOF-5 (Mg) (c) MOF-5 (Ca) (d) MOF-508.

So far, only phonons at the Γ point were compared, but for heat transport the phonon dispersion is crucial. Figure S7 shows a comparison between the MOF-FF calculated phonon band structure for MOF-5 (Zn) and previously reported ab initio data.[24] The MOF-FF phonon band structures were obtained by using finite displacements in a 3×3×3 primitive unit cell of MOF-5 with the phonopy code.[14] The shape of most bands obtained by MOF-FF is reasonably similar to the reference, but some modes appear a bit higher or lower in energy. The crucial part of this particular comparison is the range of low-frequency modes. Especially important is the slope of the acoustic modes originating at Γ, which match very well between the MOF-FF results and the reference data in most directions. One exception is the Γ-W direction, where the agreement is slightly less satisfactory, but even at the W-point, the difference in wavenumbers amount to only 4.3 cm$^{-1}$. Thus, the overall agreement with the PBE reference data is excellent for the acoustic phonons and also properly describes a large majority of the low frequency optical phonons.

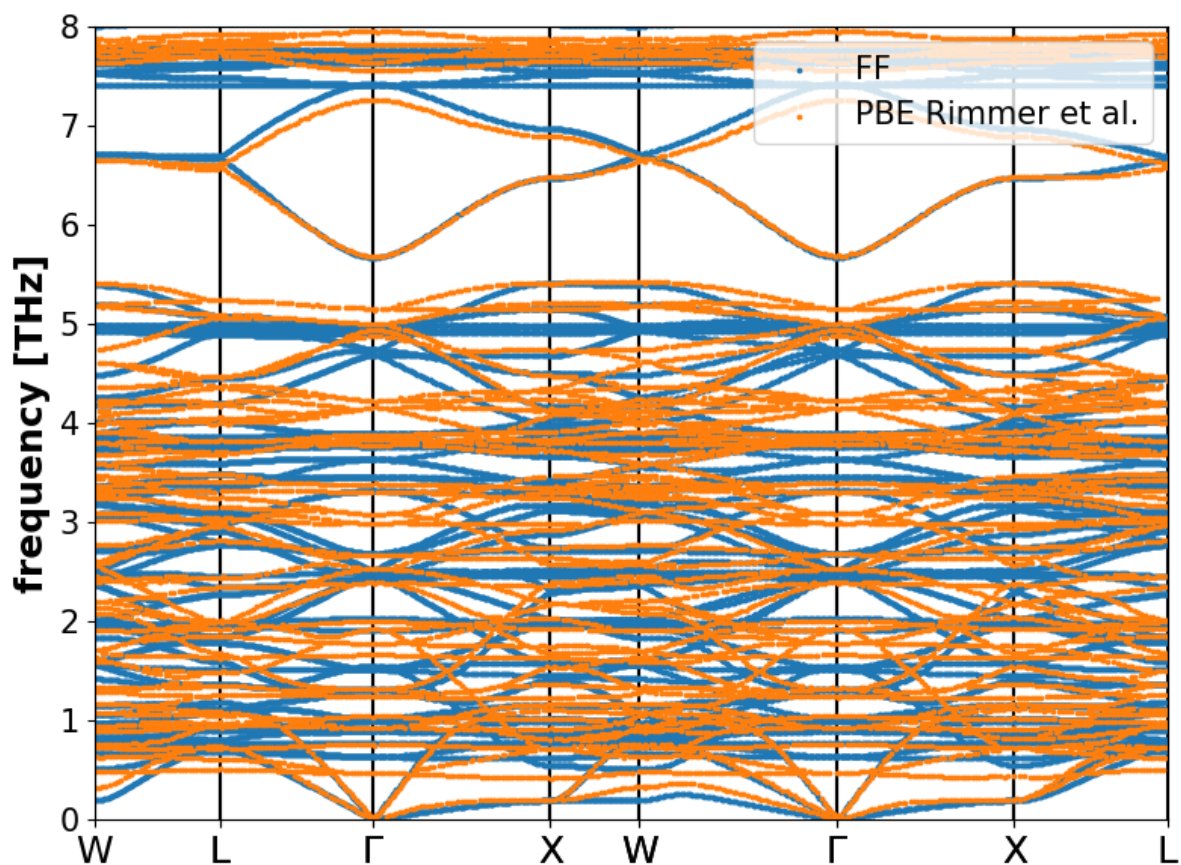

**Figure S7.** Comparison of phonon bands of the resulting force field for MOF-5 (Zn) compared to ab-initio reference extracted by the work from Rimmer et al.[24]. The ab-initio dispersion curves were obtained employing density-functional perturbation theory with the GGA-PBE functional[7] in the CASTEP program[25] and dispersion corrections according to the Grimme-D2 scheme.[26]

As a final comparison for harmonic phonon properties, we compared the MOF-FF phonon frequencies to previously reported experimental inelastic neutron scattering (INS) spectra for MOF-5 at a temperature of 20 K[27] (Figure S8). The obtained spectrum technically encompasses more than just frequencies at the Γ-point, but for clarity, only these are shown. It can be seen that the frequencies obtained with MOF-FF agree well with the low-frequency peaks in the INS spectrum (to within about 3 cm$^{-1}$).

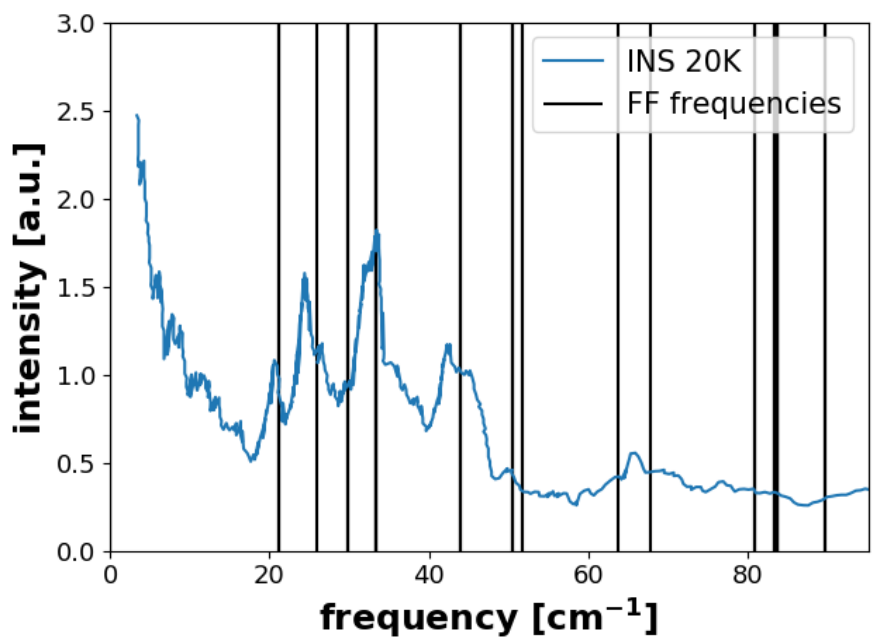

**Figure S8.** Inelastic neutron scattering spectrum of MOF-5 (Zn) at a temperature of 20 K compared to the Γ-point phonon frequencies of MOF-5 (Zn), calculated with the MOF-FF force field parametrized here. Experimental spectrum extracted from the work of Lock et al..[27]

### S4.3  Thermal expansion

In addition to the just discussed harmonic phonon properties, anharmonic properties play an important role in heat transport. Anharmonic effects for each phonon mode are currently unfeasible to calculate using high-level DFT simulations. Therefore, we use the experimental reported thermal expansion, a property caused by anharmonic effects, as an indicator of the quality of the force field regarding anharmonic properties. Contrary to most other materials, the thermal expansion of many MOFs is known to be negative[28,29]. For MOF-5 (Zn), several measurements have been carried out to analyze this unusual behavior. The experimental values for the thermal expansion coefficient range from -13.1 10$^{-6}$ K$^{-1}$[29] to -16 10$^{-6}$ K$^{-1}$.[27,28] With our parametrization of MOF-FF for MOF-5 (Zn), a negative thermal expansion coefficient of -15.55 $10^{-6}$ $K^{-1}$ is obtained, which very favorably compares to the experimental values. The numeric value was calculated from a series of temperature-dependent volumes obtained from molecular dynamics simulations. The systems were equilibrated for a duration of 1.5 ns at a timestep of 0.5 fs in an isobaric-isothermal ensemble (NPT), and the resulting lattice parameters were obtained by averaging their values over the duration of the simulation after the steady state had been reached (after 10 ps). Then a linear fit to all data points was performed to obtain the thermal expansion coefficient. The data points and the fit are visualized in Figure S9.

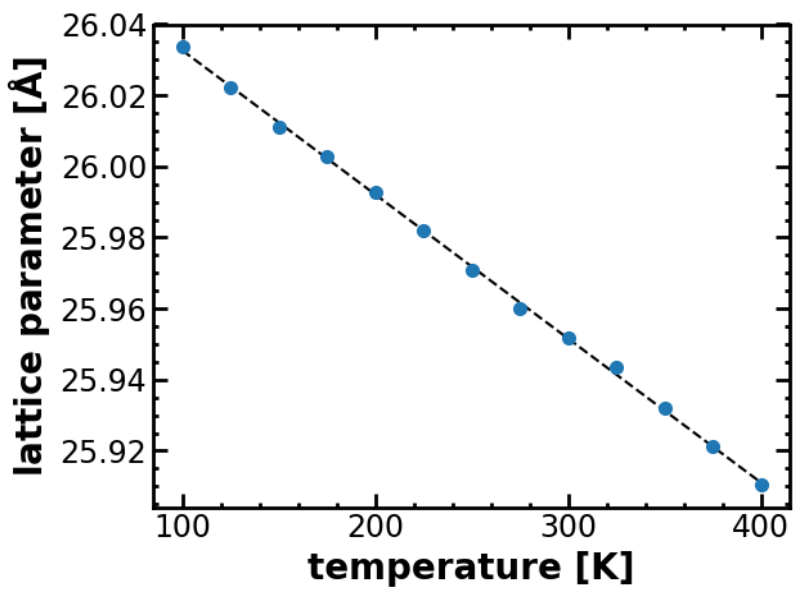

**Figure S9.** Lattice parameter obtained by NPT simulations of MOF-5 (Zn) as a function of temperature showing linear negative thermal expansion.

## S5. Details of the molecular dynamics simulations

Molecular dynamics represents an attractive method for obtaining the thermal conductivity of a material due to the inherent inclusion of anharmonic effects in a non-approximative way (beyond the inaccuracies of the way forces are calculated). It is possible to carry out thermal conductivity simulations at equilibrium and non-equilibrium. We employ the non-equilibrium molecular dynamics (NEMD)[30] approach, as we intend to spatially resolve temperature profiles. There, one builds a supercell of the material, which is usually longer along the direction into which the thermal conductivity is calculated. Two small equally sized sections along the cell length are defined as the heat source and sink. First, the entire simulation box is heated at the equilibrium temperature for which the thermal conductivity will be investigated. Heat is then introduced through the heat source and removed from the heat sink. This results in a temperature gradient $\nabla T$ and a heat flux $j$ in the remaining bulk of the system. The thermal conductivity can be obtained using Fourier's law.

$$j = -\kappa \nabla T \qquad (S2)$$

In the resulting temperature profiles, a typical temperature drop can be observed in proximity to the thermostat, as can be seen in Figure S10. This is a consequence of finite size effects, when the distance between the thermostats is not significantly longer than the maximum phonon mean free path. As a result, phonon scattering at the thermostat boundary occurs,[30] which reduces the thermal conductivity. To properly obtain a representative numerical value, multiple NEMD simulations with different cell lengths are required. This allows the extrapolation of the thermal conductivity towards the infinite size limit using a linear fit of $1/\kappa$ over $1/L$,[30,31] as represented in Figure S11.

All molecular dynamics simulations were carried out with the LAMMPS software.[19] Long-range electrostatic interactions were included with a particle-particle particle-mesh eigensolver.[32] The simulations with the neural network potential were carried out using the n2p2 extension for LAMMPS, which implements the neural network potentials developed by Behler et al..[33] Input files used for the molecular dynamics simulations are also included in the accompanying repository at https://github.com/sandrowieser/mof_thermal_bottleneck_si.

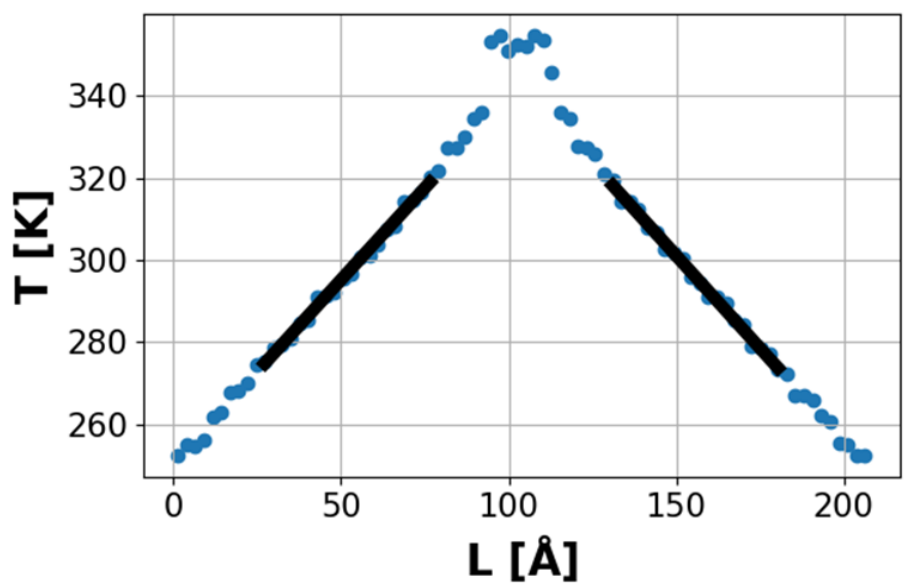

**Figure S10.** Temperature profile of an 8×1×1 cell of MOF-5 (Zn) during a NEMD simulation in heat flux direction. The straight line indicates the region of the linear fit.

### S5.1 Determination of lattice parameters at 300 K

Before performing the thermal conductivity simulations, the equilibrium volume of the investigated systems at 300 K was determined to account for thermal expansion properly. A 2×2×2 conventional supercell of MOF-5 (containing 64 nodes and 192 linkers) was equilibrated in an isobaric-isothermal ensemble (NPT) for at least 500 ps with a time step of 0.5 fs. The resulting average cubic lattice parameters for different MOF variants are listed in Table S4 together with their respective value at the energy minimum at 0 K and with the DFT reference. It can be observed that all MOF-5 values agree well with the ab initio reference data. Still, for MOF-508, there seems to be a slight discrepancy of the lattice parameter in the direction of the bipyridine linkers. All variants show a negative thermal expansion coefficient, which has previously been shown to be the case for MOF-5.[27] The lattice parameter of 25.951 Å for MOF-5 (Zn) compares well to the experimental reference of 25.8496 Å at 300 K[34]. The Mg and Zn MOFs show a very similar lattice parameter, while the value for Ca is significantly higher. This is due to a difference in the metal-oxygen bond length and a generally weaker bond strength at this bond (see section S6.1).

**Table S4.** Lattice parameters in Å of the investigated MOF-5 variants and MOF-508 in directions of the 1,4-benzenedicarboxylate (BDC) and the bipyridine (bipy) linkers obtained from energy minimization at 0 K and from NPT molecular dynamics simulations at 300 K.

| System | DFT, 0 K [Å] | FF, 0 K [Å] | FF, 300 K [Å] |
|---|---|---|---|
| MOF-5 (Mg) | 26.139 | 26.139 | 26.020 |
| MOF-5 (Ca) | 27.760 | 27.760 | 27.476 |
| MOF-5 (Zn) | 26.074 | 26.074 | 25.951 |
| MOF-508 (Zn) BDC | 11.069 | 11.064 | 11.024 |
| MOF-508 (Zn) bipy | 13.997 | 14.043 | 13.991 |

### S5.2 Converging and benchmarking the NEMD simulations

For the NEMD simulations, simulation boxes of different lengths ranging from 8×1×1 to 32×1×1 conventional cubic unit cells of MOF-5 were used. The thickness of the cell of about 26x26 Å (2 nodes × 2 nodes) was deemed sufficient, as 8×1×1 and 16×1×1 cells yielded thermal conductivities that were within 0.1 W(mK)$^{-1}$ of the values obtained for the 8×2×2 and 16×2×2 cells. For MOF-508, however, cells with a thickness of 4×4 (4 nodes × 4 nodes) unit cells in the direction perpendicular to heat flux were used, since significant differences between 2×2 and 4×4 cells were observed. Therefore, for the smaller unit cells (one node and three linkers) of MOF-508, cell sizes ranging from 16×4×4 to 40×4×4 were used to perform the NEMD simulations. The values for the differences between the thermal conductivities for individual cells are listed in Table S5.

**Table S5.** Thermal conductivity comparison resulting from NEMD simulations for systems of different cell thickness of 2 nodes and 4 nodes for MOF-5 (Zn) (×1×1 cells and ×2×2 cells) and MOF-508 (Zn) (×2×2 cells and ×4×4 cells).

| System | $\kappa_{2\ nodes}$ [W(mK)$^{-1}$] | $\kappa_{4\ nodes}$ [W(mK)$^{-1}$] |
|---|---|---|
| MOF-5 (Zn) 16 nodes long | 0.241 | 0.237 |
| MOF-5 (Zn) 32 nodes long | 0.297 | 0.289 |
| MOF-508 (Zn) 16 nodes long | 0.182 | 0.306 |

For the reference simulations with the neural network potential (NNP)[35] for MOF-5, 6×2×2 to 12×2×2 cells were used.

As the first step of the NEMD simulations, each cell was equilibrated in the canonical ensemble (NVT) at 300 K for at least 50 ps to reach the desired simulation temperature. Subsequently, the global thermostat was removed, and the heat bath and sink were introduced by two Langevin thermostats with a friction coefficient of 200 fs and a temperature difference of ±50 K from equilibrium. The size of the thermostat in the heat flux direction was set to include exactly one node and a linker. The rest of the simulation box was then run in a microcanonical ensemble (NVE) until a steady state was reached. The resulting temperature profile and the heat flux averaged until convergence was reached. This was achieved confidently after 10 ns with a time step of 0.5 fs. Then, the temperature gradient was measured from the temperature profile using a linear fit of the bulk regions between the thermostats. It is important to note that close to the thermostat, scattering effects can occur, which can lead to a non-linear temperature profile. Therefore, the fits have to be performed in a suitable region between the thermostats showing the desired linear drop of the temperature with the position, as indicated in Figure S10. The data was then used to fit the thermal conductivity to the infinite size limit for the individual systems, as described above (see Figure S11). For each infinite size limit fit (including), the results of at least five different NEMD simulations have been used.

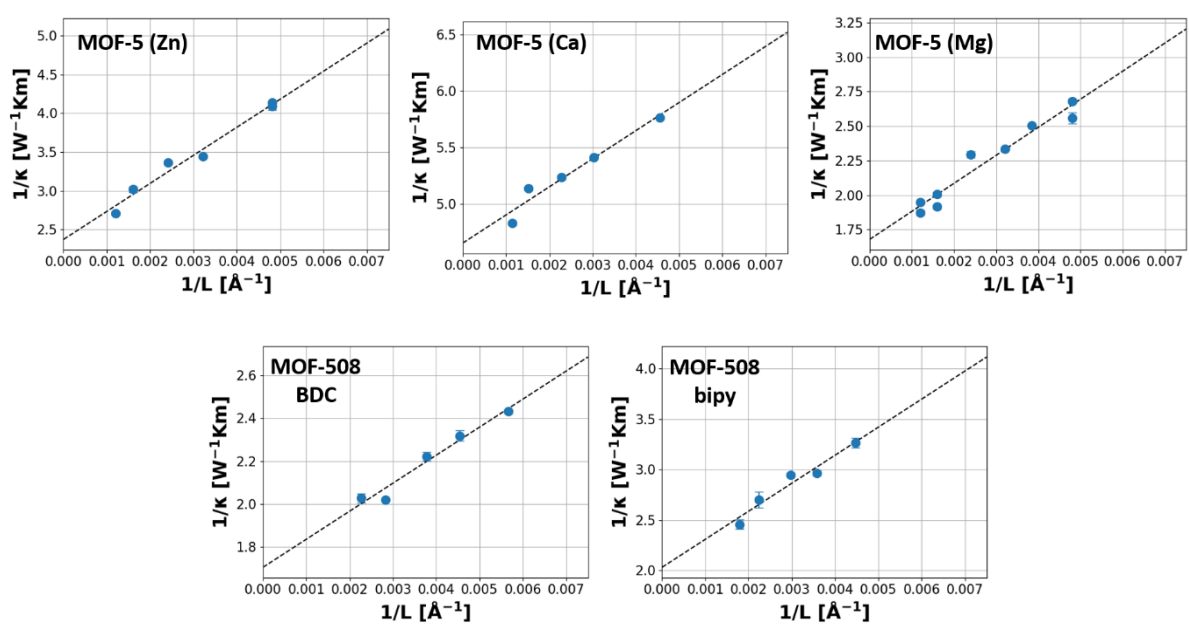

**Figure S11.** NEMD infinite size limit fits to obtain the thermal conductivity of MOF-5 (Zn), MOF-5 (Ca), MOF-5 (Mg), and MOF-508 (Zn) in BDC and bipy direction. The relation of the inverse of the thermal conductivity obtained from a NEMD simulation to the inverse of the simulated box length is visualized. A linear fit is then performed to extrapolate to the infinite size limit of the thermal conductivity by extracting the value at $1/L = 0$.

### S5.3 Evaluation of the thermal resistance contributions

To evaluate the temperature profiles for estimating the thermal resistance contributions of node, linker, and interface, we used the data from NEMD simulations of the 8×1×1 cells for MOF-5 and 16x4×4 cells for MOF-508. Smaller cell lengths resulted in larger relative temperature differences between the individual plateaus observed in the temperature profile leading to lower noise levels in the results. The full temperature profiles of all the systems used for the thermal resistance evaluation are shown in Figure S12. Additionally, the linkers perpendicular to the nodes have been neglected in the analysis, as they would lead to an artificial flattening of the temperature profile of the node. They are connected centrally to the node, and on average, have the temperature occurring at this position. This underestimates the thermal resistance of the node one would provide the false impression that exclusively the interface between node and linker is essential for thermal transport. This problem is visualized in Figure S13.

The local thermal resistance, $R_{ab}^{th}$, of a specific region of the sample between positions $z_a$ and $z_b$ multiplied by the cross-section of the thermal conductor, A, is given by:

$$R_{ab}^{th} \times A = \frac{\overline{T_a'} - \overline{T_b'}}{j} = \frac{\nabla_z \overline{T'}}{j}(z_b - z_a) = \frac{1}{\kappa_{ab}}(z_b - z_a) \tag{S3}$$

Reporting $R_{ab}^{th} \times A$ rather than $R_{ab}^{th}$ yields values that are independent of the actual cross-section of the sample and are also consistent with the definition of the Kapitza interface resistance[36]. $\overline{T_a'}$ and $\overline{T_b'}$ are the local effective temperatures of the equivalent atoms at positions $z_a$ and $z_b$, respectively. $\nabla_z \overline{T'}$ is the average gradient of $\overline{T'}$ with respect to direction z in the region between $z_a$, and $z_b$ and is determined from a linear fit to the $\overline{T'}(z)$ in that region. $j$ is the heat flow per area and $\kappa_{ab}$ the average thermal conductivity of the region. The choice of $z_a$ and $z_b$ determines, whether the thermal resistance of, e.g., a linker or a node is calculated. In the current manuscript, the boundaries between the regions ($z_a$ and $z_b$) were chosen halfway between the terminal O atoms of the linkers and the metal atoms of the nodes (indicated as dashed red lines in Figure 2a and Figure 2b in the main manuscript). The remaining temperature difference between the linear fits at these positions, $\overline{\Delta T_{int}'}$ determines the Kapitza resistance of the interface as:

$$R_{Kapitza} = R_{interface}^{th} \cdot A = \frac{\overline{\Delta T_{int}'}}{j} \tag{S4}$$

From the perspective of thermal transport, the unit of the MOF periodicity in the direction of heat flow comprises one node, one linker, and two interfaces connecting a node to the linkers on either side. Note, that in the case of MOF-5, this does not correspond to the crystallographic unit cell, which contains twice as many linkers and nodes, due to alternating torsions of the BDC linkers. We, however, did not observe a significant effect on the thermal resistance contribution for different linker torsions. The total thermal resistance per unit $R_{unit}^{th}$ can then be calculated as the sum of the individual contributions as given in Equation 1 in the main manuscript.

For MOF-508, this analysis in the direction of the BDC linkers is not possible in the same way as performed for the other systems. The reason for that is that the particular extent of the "node" in that direction corresponds to only a single atom. Therefore, instead of performing linear fits independently across nodes and linkers, here, only the temperature profile within the linker was fitted, and the remaining temperature step is associated with two interface resistances between linker and node.

The numerical values of all thermal resistance contributions are given in Table S6.

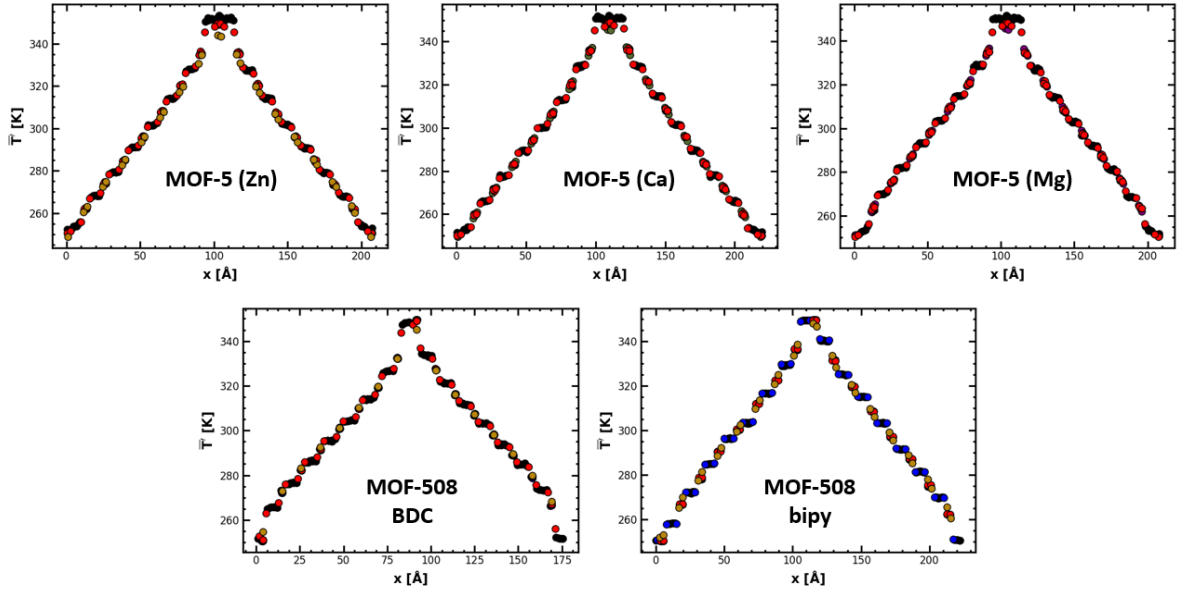

**Figure S12.** Local temperature profiles for MOF-5 with different metals and MOF-508 (Zn) in different heat flux directions for the respectively smallest simulation cells (8×1×1 for MOF-5 variants and 16×4×4 for MOF-508 each containing 16 nodes in heat-transport direction).

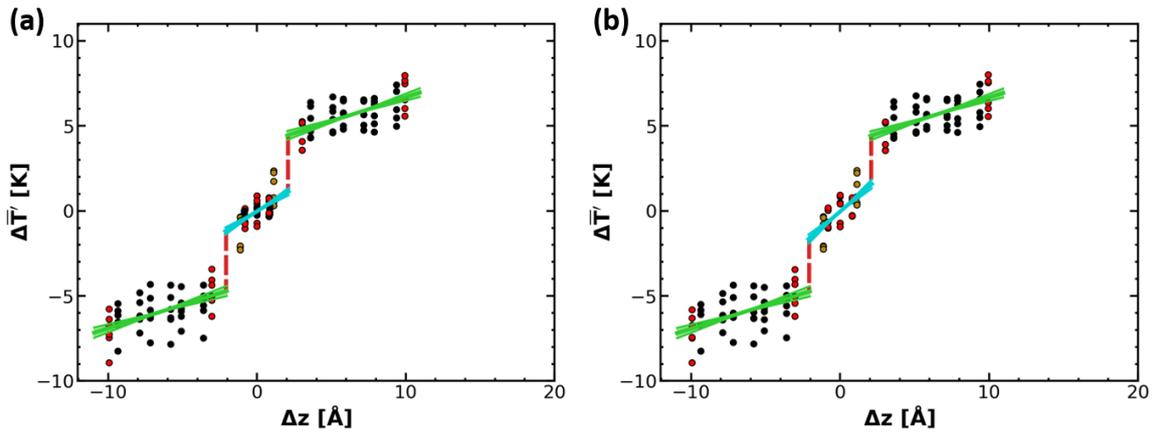

**Figure S13.** Local temperature profile of MOF-5 (Zn) during a NEMD simulation with overlaid linkers and nodes occurring in the system accounting for the cross-linkers perpendicular to heat flow at the location of the node (a) and neglecting the cross-linkers (b). The circles indicate local temperatures of the atoms (Zn, gold; O, red; C, black). Hydrogen atoms are omitted for clarity. Linear fits of individual segment contributions are indicated with the green and blue lines, while the red dashed lines represent the interface in between. Note that there is no temperature gradient for the carbon atoms forming the linkers perpendicular to the node. It can be seen that the slope through the node in (a) is lower than in (b). This is due to the carbon atoms in the linkers perpendicular to the direction of heat flow, which appear at $\Delta z$ values characteristic of the nodes, as can be seen in (a). They have essentially the same average local temperature as the central atoms in the node.

**Table S6.** Contributions to thermal node, linker and interface resistance obtained from the evaluation of the temperature profiles of MOF-5 with Mg, Ca and Zn as metal, as well as for MOF-508 in direction of BDC and bipy linkers.

| System | $R^{th}_{linker}A$ [Km$^2$GW$^{-1}$] | $R^{th}_{node}A$ [Km$^2$GW$^{-1}$] | $R^{th}_{interface}A$ [Km$^2$GW$^{-1}$] |
|---|---|---|---|
| MOF-5 (Zn) | (1.1 ± 0.3) | (1.60 ± 0.3) | (2.78 ± 0.3) |
| MOF-5 (Ca) | (1.1 ± 0.4) | (2.45 ± 0.4) | (4.45 ± 0.4) |
| MOF-5 (Mg) | (0.7 ± 0.2) | (0.99 ± 0.2) | (1.74 ± 0.2) |
| MOF-508 (Zn) BDC | (0.6 ± 0.2) | - | (1.9 ± 0.1) |
| MOF-508 (Zn) bipy | (0.1 ± 0.2) | (1.0 ± 0.2) | (3.6 ± 0.3) |

## S6. Bond lengths and bond force constants for MOF-5 (Zn), MOF-5 (Ca) and MOF-5 (Mg)

Table S7 and Table S8 compares bond lengths and bond strengths for the individual bonds in the Zn, Ca, and Mg variants of MOF-5. The bond lengths were taken from the optimized DFT reference geometries (Table S7) and from the force-field data (Table S8). The bond coupling parameters were calculated from the force constants after a transformation in internal coordinates. The values from DFT and from the FF are almost identical. It can be seen that Mg and Zn are very similar, while Ca shows significantly longer and weaker bonds within the node.

**Table S7.** Bond lengths $L_{bond}$ and force constants for bond stretching $FC_{bond}$ for MOF-5 variants obtained from the DFT reference. Lengths are given in Å and force constants in kcal(Å²mol)⁻¹

| Interaction | MOF-5 (Zn) | | MOF-5 (Ca) | | MOF-5 (Mg) | |
| --- | --- | --- | --- | --- | --- | --- |
| | $L_{bond}$ | $FC_{bond}$ | $L_{bond}$ | $FC_{bond}$ | $L_{bond}$ | $FC_{bond}$ |
| metal-$O_{cen}$ | 1.9648 | 180.9 | 2.2717 | 134.7 | 1.9869 | 170.0 |
| metal-$O_{co2}$ | 1.9670 | 188.4 | 2.2470 | 144.6 | 1.9602 | 189.3 |
| $O_{co2}$-$C_{co2}$ | 1.2769 | 818.4 | 1.2763 | 843.6 | 1.2760 | 829.8 |
| $C_{co2}$-$C_{c3}$ | 1.4960 | 525.8 | 1.4979 | 516.8 | 1.4937 | 530.0 |
| $C_{c3}$-$C_{ph}$ | 1.4026 | 676.8 | 1.4027 | 675.9 | 1.4027 | 676.9 |
| $C_{ph}$-$C_{ph}$ | 1.3904 | 674.8 | 1.3908 | 673.1 | 1.3903 | 674.7 |
| $C_{ph}$-H | 1.0887 | 776.8 | 1.0891 | 774.8 | 1.0889 | 775.9 |

**Table S8.** Bond lengths $L_{bond}$ and force constants for bond stretching $FC_{bond}$ for MOF-5 variants obtained from the respective force fields. Lengths are given in Å and force constants in kcal Å⁻²mol⁻¹

| Interaction | MOF-5 (Zn) | | MOF-5 (Ca) | | MOF-5 (Mg) | |
| --- | --- | --- | --- | --- | --- | --- |
| | $L_{bond}$ | $FC_{bond}$ | $L_{bond}$ | $FC_{bond}$ | $L_{bond}$ | $FC_{bond}$ |
| metal-$O_{cen}$ | 1.9648 | 180.8 | 2.2716 | 134.5 | 1.9868 | 169.9 |
| metal-$O_{co2}$ | 1.9668 | 188.2 | 2.2470 | 144.4 | 1.9600 | 189.2 |
| $O_{co2}$-$C_{co2}$ | 1.2771 | 818.5 | 1.2764 | 843.5 | 1.2760 | 829.8 |
| $C_{co2}$-$C_{c3}$ | 1.4960 | 525.8 | 1.4978 | 516.8 | 1.4937 | 530.0 |
| $C_{c3}$-$C_{ph}$ | 1.4026 | 676.8 | 1.4027 | 675.9 | 1.4027 | 676.9 |
| $C_{ph}$-$C_{ph}$ | 1.3904 | 674.8 | 1.3907 | 673.1 | 1.3904 | 674.7 |
| $C_{ph}$-H | 1.0887 | 776.8 | 1.0891 | 774.8 | 1.0889 | 775.9 |

### S6.1 Impact of the Ca-O coupling strength on the thermal conductivity

For rescaling the Ca-O bond strength, the bond-stretching force field parameter for the bonds of the Ca atom with the central and outer oxygen atoms have been doubled and quadrupled. The interactions are indicated in the parameter files (which can be found in the accompanying repository at https://github.com/sandrowieser/mof_thermal_bottleneck_si) in the list of `bond_coeff` as

```
mm3->(ca4_o4@ca4o,o4_ca4@ca4o)|mof5ca
```
and
```
mm3->(ca4_o4@ca4o,o2_c1ca1@co2)|mof5ca.
```

When analyzing the forces in internal coordinates, this results in coupling constants increasing by 70 and 200 %, respectively. The change is less significant here since there are additional interactions included in the force field that also impact the force constants (e.g. electrostatic interactions, which are very substantial for the Zn and O atoms showing large positive and negative atomic charges).

Performing the thermal conductivity simulations for the modified systems leads to thermal conductivity values of 0.28 for a 70 % increase of the coupling constant and 0.36 W(mK)$^{-1}$ for a 200 % increase after correcting for finite-size effects. This is higher compared to the value of 0.22 W(mK)$^{-1}$ for base MOF-5 (Ca). Additionally, the thermal resistances, which are visualized in Figure S14, do show the expected trend of the reduced interface contributions. The increase in thermal conductivity originates from a significant reduction of the interface thermal resistance. Note, that the total thermal conductivity is still not as high as for MOF-5 (Zn) with the same metal mass as for Ca indicating that there must be another effect influencing the thermal conductivity. We attribute this to the larger bond length between Zn and O, which is not modified by the change of the coupling constant. Unfortunately, this parameter cannot be manipulated easily in the force field, without having a major impact many other interactions occurring in the system. Especially cross-terms describing interactions between individual bonds, which also depend on the lengths of the bond would be affected. Thus, it is impossible to analyze that contribution individually.

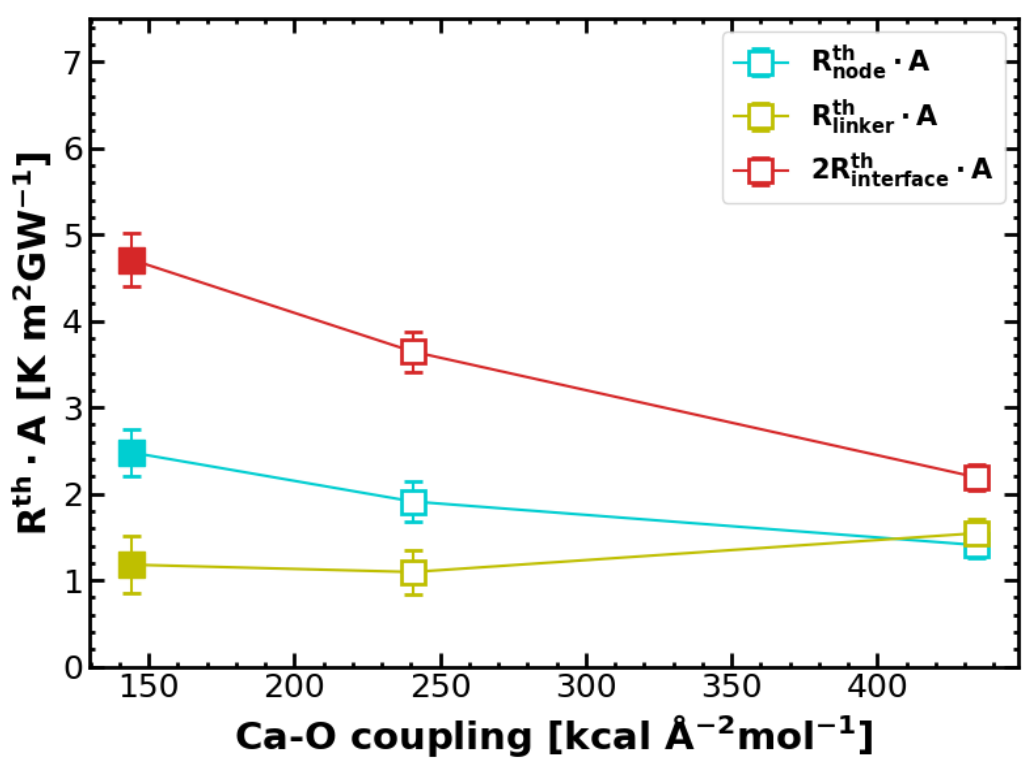

**Figure S14.** Thermal resistance contributions for MOF-5 (Ca) with modified Ca-O coupling force constants. The filled symbols indicate the original system.